\documentclass[11pt,a4paper]{article}
\usepackage{jheppub}
\usepackage{graphicx}
\usepackage{dcolumn}

\usepackage{amsmath}
\usepackage{bm}
\usepackage{yhmath}
\usepackage{textcomp}

\usepackage{color}
\usepackage{ulem}

\allowdisplaybreaks

\title{A Renormalizable Supersymmetric SU(5) Model}
\author[]{Jun-hui Zheng,}
\author[]{and Da-Xin Zhang}
\affiliation[]{School of Physics
and State Key Laboratory of Nuclear Physics and Technology,\\
Peking University, Beijing 100871, China}
\emailAdd{10904013@pku.edu.cn}
\emailAdd{dxzhang@pku.edu.cn}
\abstract{In the Supersymmetric SU(5) Model of Unification with the  Missing
Partner Mechanism,  we present a renormalizable model using the
Georgi-Jarlsog mechanism to describe the fermion masses and mixing.
At the meantime the proton decay rates are also suppressed to
satisfy the experimental data.}

\begin{document}

\maketitle

\pagenumbering{arabic}

\section{Introduction}\label{sec1}

A Grand Unified Theory (GUT)  model need to realize the unification
of the gauge couplings and give the correct fermion masses and
mixing. In the SU(5) Supersymmtric (SUSY) GUT (SGUT) models, the
unification of the gauge couplings \cite{luo1991,ellis1991} can be
achieved by taking threshold effects into account. The correct
fermion masses and mixing, however, cannot be given in the minimal
version of the SU(5) model which contains only $5+\bar 5+24$ in the
Higgs sector. In addition, the threshold effects in realizing the
gauge coupling unification constrain the  spectrum of the entire
model\cite{murayama1993}. In the minimal version of the SU(5) model,
these constraints are quite strong. The resulting color-triplet
Higgsino masses of the $5+\bar 5$ are rather low, which induce too
rapid proton decay rates to be acceptable
experimentally\cite{murayama2002}. A related issue is the
doublet-triplet splitting problem which requires a pair of nearly
massless weak doublets  to break the electroweak symmetry.

There are many efforts to solve the above problems.
 To give the correct fermion masses and mixing in the
non-SUSY SU(5) model, the Georgi-Jarlskog mechanism
(GJM)\cite{georgi1979} can be used by introducing extra Higgs of
$\overline{45}$. In the SUSY version of the GJM, Higgs superfields
of $45+\overline{45}$ need to be added\cite{wuzhang}. Proton decay
can be suppressed by the cancelation of different color-triplet
pairs of Higgsino. However, the up-type quarks get masses not only
from the original Yukawa couplings of $5-10_F-10_F$ (the subscript
$F$ stands for fermion or matter) but also from the newly introduced
Yukawa couplings of $45-10_F-10_F$. This may induce more
unobservable parameters in the up-type Yukawa
couplings\cite{dorsner2005}, although proton decay rates can be
further suppressed since the relation is weakened between the
fermion masses and the dimension-5 operators generated by the
color-triplet Higgsinos.

People usually use the Missing Partner Mechanism
(MPM)\cite{Yanagida1982,Grinstein1982} to resolve the
doublet-triplet splitting problem.  In the MPM  a pair of
$50+\overline{50}$ Higgs are introduced which contain a pair of
color-triplet Higgs without new weak-doublet Higgs. Instead of $24$,
a $75$-Higgs is used to break the SU(5) symmetry. By introducing a
U(1) symmetry\cite{alteralli2000}, the suitable superpotential can
be written to generate a pair of massless weak doublets. All Higgs
superfields but a pair of weak doublets are heavy so that at low
energy the model recover to be the Minimal SUSY Standard Model
(MSSM).

In the present work we will give a realistic model by applying the
GJM and the MPM simultaneously. The Higgs sector contains $1, 5+\bar
5, ~45+\overline{45}, ~50+\overline{50}, ~75$ multiplets of SU(5),
while the matter sector remains the same as in the minimal model.
All the couplings are renormalizable in the present model, unlike in
\cite{alteralli2000} where high-order couplings are used to generate
fermion masses through the Frogatt-Nielsen mechanism\cite{FNM}. By
introducing two $U(1)$ symmetries and giving the $45$  different
U(1) charges from the $5$'s, the up-type quarks get masses from the
$5$ only. This eliminates the extra couplings in \cite{wuzhang} and
makes the model rather predictable.

We will present the model explicitly. Then we carry the analysis on
the GUT spectrum, calculate the constraints imposed by gauge
coupling unification, study the fermion masses and proton decay, and
conclude.

\section{The Model and the Spectrum}
In the GUT models with MPM to realize the doublet-triplet splitting,
extra Higgs of $50+\overline{50}$ are added. To protect the  gauge
coupling evolution  perturbatively below the GUT scale, these Higgs
need to be as heavy as around the Planck scale. This is realized by
introducing a $U_P(1)$ symmetry  breaking just below the Planck
scale. Furthermore, for the model to recover the MSSM as all but one
pair of Higgs doublets are massless at the GUT scale, another
symmetry $U_S(1)$ is added to break at a scale below the GUT scale.
The Higgs sector contain  two singlets, two pairs of $5+\bar 5$, two
pairs of $50+\overline{50}$, and one $75$ to break the SU(5). This
is in accord with the improved MPM of \cite{bere}. In addition, a
pair of $45+\overline{45}$ is needed to generate fermion masses and
mixing through the GJM. We assign the $U_S(1)$ and $U_P(1)$ charges
for the Higgs multiplets as in Table I.

\begin{table}
  \centering
  \caption{$U_S(1)$ and $U_P(1)$ quantum numbers for the Higgs superfields.}
  \label{t0}
  \begin{tabular}{|c|c|c|c|c|c|c|c|c|c|c|c|c|c|}
   \hline
   Higgs & $5$ & $\overline {5}$ & $45$ & $\overline {45}$ & $5'$ & $\overline {5'}$ & $50$ & $ \overline {50}$ & $ 50'$  & $ \overline {50'}$ & $75$ & $S$ & $P$\\
   \hline
    $U_S(1)$ & $h$ & $-q-h$ & $q+h$ & $-q-h$ & $q+h$ & $-h$ & $q+h$ & $ -h$ & $h$  & $ -q-h $ & $0$ & $-q$ & $0$\\
   \hline
     $U_P(1)$ & $\tau$ & $-\sigma$ & $\sigma$ & $-\sigma$ & $\tau$ & $-\sigma$ & $\sigma$ & $ -\tau$ & $ \sigma$  & $ -\tau$ & $0$ & $-\tau+\sigma$ & $\tau-\sigma$\\
   \hline
   \end{tabular}
\end{table}

The superpotential of Higgs sector  is
\begin{equation}\label{2}
    \begin{split}
      W= & -\frac{\sqrt{2}}{6} \lambda A 75^{ij}_{kl}75_{ij}^{kl} +\lambda 75^{ij}_{kl}75^{km}_{in}75^{nl}_{mj} \\&
      + \frac{m}{2} 45_{k}^{ij} \overline{45}^{k}_{ij}+\frac{3}{2 \sqrt{2}}e \overline{5}_i  75 ^{im}_{jk} 45^{jk}_m
      + \frac{\Delta}{\langle S \rangle} S 5'^i\overline{5'}_i\\&
      +\frac{\sqrt{3}}{4 \sqrt{2}} a 5^i 75^{jk}_{lm}\overline{50}^{lm}_{ijk} +\frac{\sqrt{3}}{4 \sqrt{2}}b \overline{5}_i 75_{jk}^{lm} 50_{lm}^{ijk}
      +\frac{\sqrt{3}}{4 \sqrt{2}} b' {5'}^{i} 75^{jk}_{lm}\overline{50'}^{lm}_{ijk} +\frac{\sqrt{3}}{4 \sqrt{2}}a' \overline{5'}_i 75_{jk}^{lm} {50'}_{lm}^{ijk}\\&
      +\frac{3 \sqrt{3}}{2 \sqrt{2}}f \overline{45}^n_{ij}75^{im}_{kl} 50^{jkl}_{nm}+ \frac{M_1}{12 \langle P \rangle} P 50 _{lm}^{ijk} \overline{50'}^{lm}_{ijk}+ \frac{M_2}{12 \langle P \rangle} P {50'} _{lm}^{ijk}
      \overline{50}^{lm}_{ijk},
    \end{split}
\end{equation}
where the coefficients are chosen for later convenience. We demand
all the trilinear couplings are of order one to avoid fine-tuning.
The mechanism of breaking the U(1)'s can be found in  e.g.
\cite{Hisano1995}. Also, we will not discuss whether these U(1) are
global \cite{alteralli2000} or anomalous\cite{bere} which are both
used in the literature.

Just below the Planck scale, the $U_P(1)$ symmetry is broken when
the SU(5) singlet $P(0,-\sigma+\tau)$ obtain a vacuum expecting
value (VEV) $\langle P \rangle$.  This leads the $50,
\overline{50},50', \overline{50'}$ to be heavy. The SU(5) symmetry
is broken when the Standard Model (SM) singlet of the $75$ obtains a
VEV $A$, while the SU(5) singlet $S(-q,\sigma-\tau)$ obtain a vacuum
value $\langle S \rangle$ to break the $U_S(1)$ symmetry at a lower
scale.

At the GUT scale the heavy Higgs $50$,$\overline{50}$, $50'$ and
$\overline{50'}$ have been integrated out, we can get the spectrum
of the Higgs multiplets in Table \ref{t1}. The spectrum is to be
constrained by the requirement of gauge coupling unification through
the threshold effects. We have neglected some small effects of the
$50$'s on the $45+\overline{45}$.

\begin{table}
  \centering
  \caption{The Higgs spectrum
  ($a,b$ are color indexes, $\alpha,\beta$ are flavor indexes.)}\label{t1}
  \begin{tabular}{|c|c|c|c|}
    \hline
    Higgs multiplets & Representation under SM group & Mass matrix & From $SU(5)$ representation \\
    \hline
    $ T^a ,\overline{T}_a $ & $(3,1,-\frac{1}{3}),(\overline{3},1,\frac{1}{3})$ & $ \left(
                                                                                  \begin{array}{ccc}
                                                                                    0 & -e A & -\frac{bb' A^2}{M_1} \\
                                                                                    0 & m & -\frac{fb' A^2}{M_1}\\
                                                                                    -\frac{a a' A^2}{M_2} & 0  & \Delta \\
                                                                                  \end{array}
                                                                                \right) $
     & $ 5+45+5'$,$ \overline{5}+\overline{45}+\overline{5'} $ \\
    \hline
    $ H^{\alpha},\overline{H}_{\alpha} $ & $ (1,2,\frac{1}{2}),(1,2,-\frac{1}{2}) $ & $ \left(
                                                                                         \begin{array}{ccc}
                                                                                           0 & \sqrt{3} e A & 0\\
                                                                                           0 & m  & 0\\
                                                                                           0 & 0  & \Delta \\
                                                                                         \end{array}
                                                                                       \right) $
     & $5+45+5'$,$\overline{5}+\overline{45}+\overline{5'} $ \\
    \hline
    $ H^{a \alpha}_b , H_{a\alpha}^b $ & $ (8,2,\frac{1}{2}),(8,2,-\frac{1}{2}) $ & $ m $
    & $45$,$\overline{45}$ \\
    \hline
    $ H^{a \alpha}, H_{a \alpha} $ & $ (3,2, \frac{7}{6}),(\overline{3},2, -\frac{7}{6}) $ & $ m$
    & $45$,$\overline{45}$ \\
    \hline
    $ H_{ab(s)}, H^{ab}_{(s)} $ & $(\overline{6} ,1 ,-\frac{1}{3}),(6 ,1 , \frac{1}{3}) $ & $ m $ & $45$,$\overline{45} $ \\
    \hline
    $H^{a \alpha}_{\beta},H_{a \alpha}^{\beta}$ & $(3,3,-\frac{1}{3}),(\overline{3},3,\frac{1}{3})$ & $m$ & $45$,$\overline{45}$ \\
    \hline
    $H^a,\overline{H}_a $ & $(3,1,-\frac{4}{3}), (\overline{3},1,\frac{4}{3})$ & $m $ & $45$,$\overline{45}$ \\
    \hline
    $ \Sigma^{a}_b $ & $ (8,1,0)$ & $ -\frac{2\sqrt{2}}{3}\lambda A $ & $ 75 $ \\
    \hline
    $ \Sigma^{a \alpha}_{b \beta} $ & $ (8,3,0) $ & $-\frac{10\sqrt{2}}{3}\lambda A $ & $ 75 $ \\
    \hline
    $ \Sigma_0 $ & $(1,1,0) $ & $\frac{4\sqrt{2}}{3}\lambda A $ & $ 75 $ \\
    \hline
    $\Sigma^{ab}_{\alpha(s)}, \Sigma_{ab(s)}^{\alpha} $ & $(6 ,2 , \frac{5}{6}),(\overline{6} ,2 ,-\frac{5}{6}) $ & $ -\frac{4\sqrt{2}}{3}\lambda A  $ & $ 75 $ \\
    \hline
    $\Sigma^a,\overline{\Sigma}_a $ & $(3,1,\frac{5}{3}),(\overline{3},1,-\frac{5}{3}) $ & $\frac{8\sqrt{2}}{3}\lambda A $ & $75$ \\
    \hline
  \end{tabular}
\end{table}

The mass matrix for the color-triplet Higgs multiplets is
\begin{equation}\label{3}
    M_T= \begin{array}{c|ccc}
             & T_5 & T_{45} & T_{5'}  \\
           \hline
           \overline{T}_{\overline{5}} & 0 & -e A & -\frac{b b' A^2}{M_1} \\
           \overline{T}_{\overline{45}} & 0 & m &  -\frac{f b' A^2}{M_1} \\
           \overline{T}_{\overline{5'}} & -\frac{a a' A^2}{M_2} & 0  &
           \Delta,
         \end{array}
\end{equation}
and that for the weak-doublets is
\begin{equation}\label{4}
    M_2= \begin{array}{c|ccc}
             & H_5 & H_{45} & H_{5'}\\
           \hline
           \overline{H}_{\overline{5}} & 0 & \sqrt{3} e A & 0 \\
           \overline{H}_{\overline{45}} & 0 & m & 0\\
            \overline{H}_{\overline{5'}} & 0 & 0 & \Delta. \\
         \end{array}
\end{equation}
Diagonalizing $M_2$ gives two pair of heavy weak-doublets with masses
\begin{equation}\label{5}
   \begin{split}
  M_+=&\sqrt{m^2+(\sqrt{3} e A)^2}, \\
  M_-=&\Delta
   \end{split}
\end{equation}
and a pair of massless weak-doublets
\begin{equation}\label{6}
    H_U=H_5, ~H_D=\overline{H}_{\overline{5}} \cos \theta - \overline{H}_{\overline{45}} \sin \theta
\end{equation}
which are the two Higgs doublets of the MSSM. Here $\cos
\theta=\frac{m}{M_+}$ and $\sin
\theta=\frac{\sqrt{3} e A}{M_+}$.

\section{Unification and Threshold Effects}
In the GUT models the spectra are constrained by the threshold
effects. In the present model, by requiring the three gauge
couplings to unify at a scale $\Lambda_{GUT}$ at 1-loop, we have
\begin{equation}\label{7}
     \begin{split}
       (3\alpha_2^{-1}-2\alpha_3^{-1}-\alpha_1^{-1})(m_z)= & \frac{1}{2 \pi}\{-2\ln{\frac{m_{SUSY}}{m_z}} \\
       & +\frac{6}{5}\ln{\frac {|\det{M_T}|^2M_{H^{ab}_{(s)}}^9  M_{H^{a\alpha}_b}^4 M_{H^{a\alpha}}^4 M_{H^a}^7  M_{\Sigma^a_b}^5M_{\Sigma^{ab}_{\alpha(s)}}^{10} M_{\Sigma^a}^{10} }{m_z^2 M_+^2M_-^2 M_{H^{a\alpha}_{\beta}}^{24} M_{\Sigma^{a\alpha}_{b\beta}}^{25}}}\} \\
       (5\alpha_1^{-1}-3\alpha_2^{-1}-2\alpha_3^{-1})(m_z)= & \frac{1}{2 \pi}\{8\ln{\frac{m_{SUSY}}{m_z}}+ 6\ln{\frac {m_V^4 M_{H^{ab}_{(s)}}M_{H^{a\alpha}_b}^4 M_{H^{a\alpha}_{\beta}}^6 M_{\Sigma^a_b} M_{\Sigma^{a\alpha}_{b\beta}}^{11}}{m_z^6M_{H^{a\alpha}}^6   M_{H^a}^5  M_{\Sigma^{ab}_{\alpha(s)}}^2
       M_{\Sigma^a}^8}}\},
     \end{split}
\end{equation}
where $M_V$ is the mass of the $X,Y$ gauge superfields. Numerically
we need to include the running effects at  2-loop by adding
approximately the corrections
\begin{equation}\label{8}
    \begin{split}
       \delta_1^{(2)} & = -\frac{1}{4\pi}\sum_{j=1}^3 \frac{1}{b_j} (3b_{2j}-2b_{3j}-b_{1j})\ln{\frac{\alpha_j(m_z)}{\alpha_5(\Lambda)}}, \\
       \delta_2^{(2)} & =-\frac{1}{4\pi}\sum_{j=1}^3 \frac{1}{b_j} (5b_{1j}-3 b_{2j}-2 b_{3j})\ln{\frac{\alpha_j(m_z)}{\alpha_5(\Lambda)}},
     \end{split}
\end{equation}
on the R.H.S. of (\ref{7}), respectively, where
\begin{equation}\label{9}
    b_i = \begin{pmatrix}
        \frac{33}{5} \\
        1\\
        -3
    \end{pmatrix},
    ~b_{ij} = \begin{pmatrix}
        \frac{199}{25} & \frac{27}{5} & \frac{88}{5} \\
        \frac{9}{5} & 25 & 24\\
        \frac{11}{5} & 9 & 14
    \end{pmatrix}
\end{equation}
are the $\beta$-functions of gauge couplings in the MSSM at 1- and
2-loop level, respectively. The number $\alpha_5(\Lambda)$ can
take simply its value at 1-loop level. The threshold corrections at
the two-loop level are expected to be small and are thus omitted.
Thus we have
\begin{equation}\label{10}
    \begin{split}
     &(3\alpha_2^{-1}-2\alpha_3^{-1}-\alpha_1^{-1})(m_z)=\frac{1}{2 \pi}\{-2\ln{\frac{m_{SUSY}}{m_z}}
     +\frac{12}{5}\ln{\frac {x|\det{M_T}|}{m_z M_+ M_-}}
     -\frac{1}{2}\sum_{j=1}^3 \frac{1}{b_j} (3 b_{2j}-2 b_{3j}-b_{1j})\ln{\frac{\alpha_j(m_z)}{\alpha_5(\Lambda)}}\},\\
     &(5\alpha_1^{-1}-3\alpha_2^{-1}-2\alpha_3^{-1})(m_z)=\frac{1}{2 \pi}\{ 8\ln{\frac{m_{SUSY}}{m_z}}
     + 12\ln{\frac {y m_V^2 M_{\Sigma}}{m_z^3 }}
     -\frac{1}{2}\sum_{j=1}^3 \frac{1}{b_j} (5 b_{1j}-3 b_{2j}-2
     b_{3j})\ln{\frac{\alpha_j(m_z)}{\alpha_5(\Lambda)}}\},
    \end{split}
\end{equation}
where $M_{\Sigma}=\frac{10\sqrt{2}}{3}\lambda A$, while $x \sim
0.00006 $ and $y \sim 2.73 $ measure the mass splitting in $75$ (see
Table II).

 The effect of the mass splitting at SUSY scale can be taken into account by replacing $\ln{\frac{m_{SUSY}}{m_z}}$
in eqs.(\ref{10}) by \cite{murayama1993}
\begin{equation}\label{10.1}
    \begin{split}
         -2\ln{\frac{m_{SUSY}}{m_z}} & \to 4 \ln{\frac{m_{\tilde{g}}}{m_{\tilde{w}}}}+ \frac{3}{5}\ln{\frac{m^3_{\tilde{u}^c}m^2_{\tilde{d}^c}m_{\tilde{e}^c}}{m^4_{\tilde{Q}}m^2_{\tilde{L}}}} -\frac{8}{5}\ln{\frac{m_{\tilde{h}}}{m_z}} -\frac{2}{5}\ln{\frac{m_H}{m_z}},  \\
        8\ln{\frac{m_{SUSY}}{m_z}} & \to  4\ln{\frac{m_{\tilde{g}}}{m_z}}+4\ln{\frac{m_{\tilde{w}}}{m_z}}+ 3 \ln{\frac{m^2_{\tilde{Q}}}{m_{\tilde{u}^c}m_{\tilde{e}^c}}}.\\
    \end{split}
\end{equation}
The parameters can be also found in \cite{murayama1993}. The
difference between $\overline{MS}$-scheme and $\overline{DR}$-scheme
can be
 found in Ref.\cite{Matin1993}. Combining with the effect of top quark \cite{Yamada1993}, we use the following
 formulas to replace  $\frac{1}{\alpha_i}$ by
 \begin{equation}\label{10.2}
    \frac{1}{\alpha_i} \to \frac{1}{\alpha_i}-\frac{C_i}{12 \pi} + D_i\ln{\frac{m_t}{m_z}},
 \end{equation}
 where $C_1=0,C_2=2,C_3=3$ and $D_1= \frac{8 \cos^2{\theta_w}}{15 \pi} ,D_2=\frac{8 \sin^2{\theta_w}}{9 \pi},D_3=\frac{1}{3 \pi} $.

 Comparing the threshold effects with those in the MSGUT, we find  an
effective color-triplet Higgs mass
\begin{equation}\label{11a}
M_{H_c}=\frac {|\det{M_T}|}{M_+ M_-},
\end{equation}
and an effective GUT scale
\begin{equation}\label{11b}
\Lambda_{GUT}=[m_V^2 M_{\Sigma}]^{\frac{1}{3}}.
\end{equation}
Together with the extra $x$ and $y$ factors, they are constrained by
\cite{murayama2002}
\begin{equation}\label{11}
    \begin{split}
     & 5.8 \times 10^{18} GeV  \leq M_{H_c} \leq 6.0 \times 10^{19} GeV,\\
     & 1.2 \times 10^{16} GeV  \leq \Lambda_{GUT}  \leq 1.4 \times 10^{16} GeV,
    \end{split}
\end{equation}
for the wino mass $ m_{\tilde{w}}=200 GeV,\alpha_3(m_z)=0.1185 \pm 0.002, \sin{\theta_w}(m_z)=0.23117 \pm 0.00016$ and $\alpha^{-1}(m_z)=127.943 \pm 0.027$. The uncertainties come
mainly from the strong coupling constant. Note that a small $x$ will
enhance $M_{H_c}$ to suppress proton decay.

The bound on $M_{H_c}$ is about an order of magnitude larger than
that in the model of \cite{alteralli2000}, where the effective mass
$m_T= \frac{m_{T_1}m_{T_2}}{m_{\phi}}$ and $m_{T_1}, m_{T_2},
m_{\phi}$ are about $10^{16}\sim 10^{17}GeV$  which constrains $m_T$
to be $10^{15}\sim 10^{18}GeV$. In the present model with the
$5'+\overline 5'$ introduced, this constraint is release by
(\ref{11a}) so that the effective color-triplet Higgs can be large.

The doublet-triplet splitting can be found in Table \ref{t3} for a
set of representative parameters. Note that the existence of a pair
of weak-doublets at $1.0\times 10^{12} GeV$ implies that this is the
$U_S(1)$ breaking scale, if we take $\frac{\Delta}{\langle S
\rangle}\sim O(1)$ which is required by avoiding  fine-tuning the
couplings in (\ref{2}).

\begin{table}
  \centering
  \caption{The Higgs, especially the doublet-triplet, spectrum at the GUT scale for $aA=a'A=bA=b'A=fA= 5 \times 10^{16} GeV $,
  $eA=10^{16}GeV$, $M_1=M_2= 10^{18}GeV$ and $M_{H_c}=5.8\times 10^{18}GeV$. }\label{t3}
  \begin{tabular}{|c|c|}\hline
    Higgs multiplets & Masses \\
    \hline
    $T^a ,\overline{T}_a$ &  $(0.25\times 10^{16} GeV , ~~0.35\times 10^{16} GeV,
    ~~1.41 \times 10^{16} GeV )$\\
    \hline
    $ H^{\alpha},\overline{H}_{\alpha} $ & $ (0, ~~1.0\times 10^{12} GeV, ~~2\times 10^{16} GeV)$\\
    \hline
    other Higgs from $45,\overline{45}$& $1\times 10^{16} GeV$\\
    \hline
  \end{tabular}
\end{table}

\section{Fermion Masses and Proton Decays}
Without introducing extra particles, the matter fields are only the
10-plets $\psi$'s and $\bar 5$-plets $\phi$'s. Their $U_S(1)$ and $U_P(1)$quantum
numbers are $(-\frac{h}{2},-\frac{\tau}{2})$ and $(q+\frac{3}{2}h,\sigma+\frac{1}{2}\tau)$, respectively.
The superpotential for the matter and the Higgs couplings is :
\begin{equation}\label{14}
    W_F=\sqrt{2} f_1^{ij} \psi_i^{\alpha\beta} \phi_{j\alpha}\overline{5}_\beta+\sqrt{2} f_2^{ij} \psi_i^{\alpha\beta} \phi_{j\gamma}\overline{45}_{\alpha\beta}^\gamma+
    \frac{1}{4}h^{ij}\epsilon_{\alpha\beta\gamma\delta\epsilon}\psi_i^{\alpha\beta}\psi_i^{\gamma\delta}5^\epsilon,
\end{equation}
where $i,j$ are generation indices.  As we have discussed in the
Introduction, the couplings of $45$-plet Higgs  and matter field are
forbidden by the $U(1)$ symmetry. Denoting $\psi_i\ni Q_i^{'}+
u_i^{'c}+e_i^{'c}$ and $\phi_i\ni d_i^{'c}+ L_i^{'}$, we have
\begin{equation}\label{15}
    \begin{split}
      W_F \supset & Q'_id_j^{'c}(f_1^{ij}\overline{H}_{\overline{5}}+\frac{1}{\sqrt{3}}f_2^{ij}\overline{H}_{\overline{45}})
      +e_i^{'c}L'_j(f_1^{ij}\overline{H}_{\overline{5}}-\sqrt{3}f_2^{ij}\overline{H}_{\overline{45}})+ h^{ij} Q'_iu_j^{'c}H_5 \\&+u_i^{'c}d_j^{'c}(f_1^{ij}\overline{T}_{\overline{5}}+f_2^{ij}\overline{T}_{\overline{45}})
      +Q'_iL'_j(-f_1^{ij}\overline{T}_{\overline{5}}+f_2^{ij}\overline{T}_{\overline{45}})-\frac{1}{2} h^{ij} Q'_iQ'_jT_5+h^{ij} u_i^{'c}e_j^{'c}T_5
    \end{split}
\end{equation}
The couplings are related to the Yukawa couplings by
\begin{equation}\label{18}
    \begin{split}
      &h=Y_u,\\
      &f_1=\frac{1}{4 \cos \theta}(3Y_d+Y_e),\\
      &f_2=\frac{\sqrt{3}}{4\sin \theta}(-Y_d+Y_e).
    \end{split}
\end{equation}
Diagonalizing $Y$'s gives
\begin{equation}\label{19}
    \begin{split}
      &Q_i^{'}=(u_i,V_{ij}d_j)^T, ~u_i^{'c}=e^{-i\varphi_i} u_i^c , ~e_i^{'c}=V_{ij}e_j^c \\
      &d_i^{'c} =d_i^c, ~L_i^{'} = L_j= (\nu_j,e_j)^T,
    \end{split}
\end{equation}
and
\begin{equation}\label{17}
    \begin{split}
      &Y_u^{ij}=h^ie^{i\varphi_i}\delta^{ij}=e^{i\varphi_i}\delta^{ij}\frac{m_{u_i}}{v_u},\\
      &Y_d^{ij}=V_{ij}^*\frac{m_{d_j}}{v_d},\\
      &Y_e^{ij}=V_{ij}^*\frac{m_{e_j}}{v_d},
    \end{split}
\end{equation}
where the mass parameters on the R.H.S. are mass eigenvalues, and
$V$ are the CKM matrix. Two of the three phases $\varphi_i$'s are
independent\cite{murayama1993}.

In the new basics, we have
\begin{equation}\label{20}
    \begin{split}
    W_F \supset &m_u^i u_iu_i^{c}+ m_d^i d_id_i^{c} +m_e^i e_i^{c}e_i+e^{-i\varphi_i}u_i^{c}d_j^{c}(f_1^{ij}\overline{T}_1+f_2^{ij}\overline{T}_2)
    +e^{-i\varphi_i}h^{ij}V_{jk}u_i^{c}e_k^{c}T_1\\&
    +Q_iL_j (-f_1^{ij}\overline{T}_1+f_2^{ij}\overline{T}_2)
    -\frac{1}{2} h^{ij}Q_iQ_jT_1.
    \end{split}
\end{equation}
The dimension-5 operators are
\begin{equation}\label{20a}
W_5=C_{ijkl}(Q_iQ_j)(Q_kL_l)+ ~D_{ijkl}(u^c_i e^c_j)(u^c_k d^c_l),
\end{equation}
where
\begin{equation}\label{21}
    C_{ijkl}=\frac{1}{2}h^{ij}[f_1^{kl}(M_T^{-1})_{11}-f_2^{kl}(M_T^{-1})_{12}]
\end{equation}
for the LLLL  operators, and
\begin{equation}
    D_{ijkl}=h^{im}V_{mj}e^{-(\varphi_i+\varphi_k)}[f_1^{kl}(M_T^{-1})_{11}+f_2^{kl}(M_T^{-1})_{12}]
\end{equation}
for the RRRR operators. Note that
\begin{equation}\label{23}
    \begin{split}
      (M_T^{-1})_{11}=&\frac{m \Delta}{\det{M_T}}\\
      (M_T^{-1})_{12}=&\frac{\Delta}{\det{M_T}} e A,
    \end{split}
\end{equation}
we have
\begin{equation}\label{24}
    \begin{split}
      & C_{ijkl} = \frac{1}{ 2 M_{H_c}}Y_u^{ij} Y_d^{kl}\\
      & D_{ijkl} = \frac{e^{-(\varphi_i+\varphi_k)}}{ 2 M_{H_c}}Y_u^{im}V_{mj} (Y_d^{kl}+ Y_e^{kl}).
    \end{split}
\end{equation}
The operators will be dressed by the charginos to form the 4-fermion
operators to calculate proton decay, as were usually done in the
literature\cite{murayama1993}.

From equation (\ref{24}), it is the effective Higgsino mass
$M_{H_C}$ which determines the coefficients of the dimensional-5
operators. A small $x$ can enhance $M_{H_C}$ through the threshold
effects (\ref{10}) and thus suppress the proton decay rates. This
coincides with the observation from (\ref{3}) that in a limit of
vanishing $\Delta$  no proton decay can be driven by the
dimensional-5 operators.

Numerically we take the Yukawa couplings $h$ and $f_{1,2}$ to
fulfill the data of fermion masses and mixing. The difference
between   $D_{ijkl}$ in the minimal SU(5) model and that in our
model is that $m_{d_i}$ is now replaced by
$\frac{m_{d_i}+m_{e_i}}{2}$.
 Following \cite{Goto1999}, the dominant mechanism for proton decay
 is through the wino dressed  LLLL-type operators for
 $p \rightarrow K^++\bar{\nu}_{\mu(e)}$ and $p \rightarrow \pi^+ +\bar{\nu}_{\mu(e)}$, and
 through the higgsino dressed  RRRR-type operators for
 $p \rightarrow K^++\bar{\nu}_{\tau}$ and $p \rightarrow \pi^+ +\bar{\nu}_{\tau}$.
We list in Table \ref{t2} the proton partial lifetimes. Note that
these partial lifetimes are generally of orders of magnitude longer
than those in \cite{alteralli2000}, as the effective color-triplet
Higgs mass in (\ref{11}) are much heavier than that in
\cite{alteralli2000}. We found that these partial
 proton lifetimes are generally enhanced and the experimental data are satisfied.
 \begin{table}
  \centering
  \caption{Proton partial lifetimes. The experimental lower limits are typically around
  $10^{32-33}$ years\cite{pdg}.}\label{t2}
  \begin{tabular}{|c|c|}
    \hline
    Decay mode & lifetime of proton \\
    \hline
    $\tau(p \rightarrow K^++\bar{\nu}_{\mu})$ & $2.3 \times 10^{35}\sim 2.5 \times 10^{37}yrs $  \\
    \hline
    $\tau(p \rightarrow K^++\bar{\nu}_e)$  & $4.7 \times 10^{36}\sim 5.0 \times 10^{38} yrs  $ \\
    \hline
    $\tau(p \rightarrow K^++\bar{\nu}_\tau)$  & $ 2.7 \times 10^{35}\sim 2.8\times 10^{37}yrs   $ \\
    \hline
    $ \tau(p \rightarrow \pi^+ +\bar{\nu}_{\mu})$  & $4.7 \times 10^{35}\sim 5.0 \times 10^{37} yrs $ \\
    \hline
    $ \tau(p \rightarrow \pi^+ +\bar{\nu}_e) $ & $9.8\times 10^{36}\sim 1.0\times 10^{39}yrs $ \\
    \hline
    $ \tau(p \rightarrow \pi^+ +\bar{\nu}_\tau) $ & $ 6.9 \times 10^{35}\sim 7.3 \times 10^{37}yrs  $ \\
    \hline
  \end{tabular}
\end{table}

\section{Summary and Discussions}
We have presented a renormalizable model of SUSY SU(5). The MPM is
used to solve the doublet-triplet splitting problem while The GJM is
used to describe the fermion masses and mixing. Two U(1) symmetries
are used. The $U_P(1)$ is broken just below the Planck scale to give
large masses to the $50$'s, so that below the GUT scale the
evolutions of the gauge couplings are perturbative. The $U_S(1)$ is
broken at a scale around $10^{12}GeV$ to enhance the  effective mass
of the color-triplet Higgs. At the meantime of describing the
correct fermion masses and mixing, the proton decay rates are also
suppressed.

 This work was supported in part by the National Natural
Science Foundation of China (NSFC) under Grant No. 10435040.

\end{document}